# Adaptation of the neural network-based IDS to new attacks detection

Przemysław Kukiełka[1], Zbigniew Kotulski[2]
[1]Research and Development Department, Polish Telecom
[2]Institute of Telecommunications, Warsaw University of Technology
{przemyslaw.kukielka@telekomunikacja.pl; zkotulsk@tele.pw.edu.pl}

**Abstract.** In this paper we report our experiment concerning new attacks detection by a neural network-based Intrusion Detection System. What is crucial for this topic is the adaptation of the neural network that is already in use to correct classification of a new "normal traffic" and of an attack representation not presented during the network training process. When it comes to the new attack it should also be easy to obtain vectors to test and to retrain the neural classifier. We describe the proposal of an algorithm and a distributed IDS architecture that could achieve the goals mentioned above.

## 1. Introduction

Because of their generalization feature, neural networks are able to work with imprecise and incomplete data. It means that they can recognize also patterns not presented during a network learning phase. That's why neural networks could be a good solution for detection of a well-known attack, which has been modified by an aggressor in order to pass through a firewall system. In that case, traditional Intrusion Detection Systems (IDS), based on the signatures of attacks or expert rules, may not be able to detect such a new version of this attack.

Unfortunately, as it could be noticed in [1] and in the results of our research (Table 2) some representation of the new attack or the normal traffic not presented during the training process cannot be properly classified by the neural network. A remedy could be adding a vector representing new traffic to the learning data set and to retrain the neural network. However, the blocking problem for such an approach is to obtain data that represent new previously not detected attacks. We presented in this paper a proposal of an algorithm and an IDS architecture that allow to collect automatically new attacks-representing data and to use them for retraining and updating weights and the number of hidden neurons in a distributed network detectors based on the neural network technology.

In our research we focus on MLP and SVM neural networks architectures. The result of the investigation is the information about the classification accuracy, represented as a number of false alarms and not detected attacks in comparison to the number of validation vectors. A new attack and a new normal traffic representation are added to the training data set in order to observe their influence on improvement of the classification process for new vectors in the tests data set.

## 2. Neural Network: a way of work

An artificial neural network is a system simulating the work of neurons in the human brain. In Fig.1 is shown a diagram of a neuron's operation.

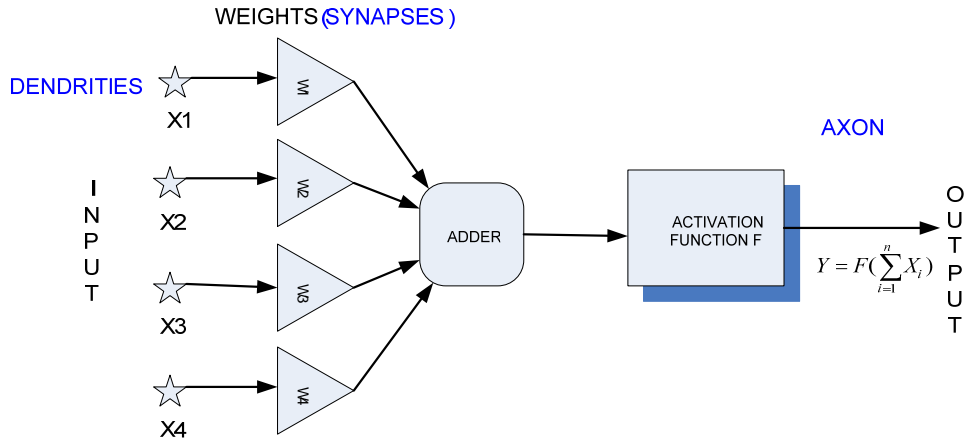

**Fig. 1 Artificial neuron's schema**

The neuron consists of a number of inputs emulating dendrites of the biological neuron, an adder module, an activation function and one output emulating an axon of the biological neuron. The importance of a particular input is indicated by its weight that emulates the biological neuron's synapse. The input signals are multiplied by the values of weights and next the results are added in the adder block. The sum is sent to the activation block where it is processed by the activation function. In that way we obtain the neuron's answer Y for the input signals "$x_i$".

One neuron cannot solve a complex problem that's why the neural network composed of many neurons is used. For the purpose of this work two network architectures were used: SVM and MLP.

One of the architectures that is used most frequently is the MLP (Multilayer Perceptron) [2], [3]. In such a network each neuron's output of the previous layer is connected with some neuron's input of the next layer. The MLP architecture consists of one or more hidden layers. The signal is transmitted through the network in one direction from the input to the output, that's why this architecture is called feedforward. The MLP network is learned with using the backward propagation algorithm (BP). In order to reach better efficiency and speed of the learning process it uses many types of the BP algorithm. In our research we used following variants of the BP algorithm:

➢ Levenberg-Marquardt (*trainln*),
➢ Resilient Backpropagation (*trainrp*).

For the simulation process Matlab toolbox was used. The variants of the BP algorithm are followed by the name of the learning Matlab's function in brackets.

The second neural network architecture that was applied for our experiment is SVM (Support Vector Machine). It is the feedforward neural network that consists of two layers (hidden and output) and can use various types of the activation function. In the classification tasks, the first step of the SVM network work is to transform the nonlinearly separated input observations (with usage of a kernel function) to the space where they can be lineally separated. The second step of the learning phase is the maximization of the separation margin between two classes of the observations. In the simulation *LibSVM* implementation of the SVM network was used.

## 3. Input data for attacks detection

In the first phase of our investigation we used the KDD 99 data set as the input vectors for training and validation of the neural networks. This data set was prepared based on the data collected in the DARPA (Defense Advanced Research Project Agency) intrusion detection evaluation program. MIT Lincoln Lab that participates in this program has set up simulation of typical LAN network in order to acquire raw TCP dump data [4]. They simulated LAN network operating as a normal environment, which was infected with various types of attacks. The raw data was processed into connection records [5], [6]. For each connection 41 various features were extracted. Each connection was labeled as the normal one or under a specific type of attack. Four main categories of the attacks were simulated:

- **DoS** (Denial of Service) – An attacker tries to prevent legitimate users from using a service, e.g. TCP SYN Flood, Smurf, etc.
- **Probe** – An attacker tries to collect information about the target host. For example: scanning victims in order to get knowledge of available services, operating system version etc.
- **U2R** (User to Root) – An attacker has a local account on the victim host and tries to gain root privileges.
- **R2L** (Remote to Local) – An attacker does not have local account on a victim host and tries to obtain it.

The KDD 99 data set is divided in to tree subsets: 10%KDD, corrected KDD, whole KDD. Basic characteristics of the KDD99 data sets are shown in Table 3. The Table 3 includes the number of connections assigned to the particular class of an attack (DoS, Probe etc.).

*Table 1. KDD 99 data subsets*

| Dataset | DoS | Probe | U2r | U2l | Normal |
|---|---|---|---|---|---|
| 10%KDD | 391458 | 4107 | 52 | 1126 | 97277 |
| Corrected KDD | 229853 | 4166 | 70 | 16347 | 60593 |
| Whole KDD | 3883370 | 41102 | 52 | 1126 | 972780 |

The 10%KDD data set is used for the training process of the IDS. It includes connections simulating 22 types of attacks and the normal traffic. The Corrected KDD data set is used for the testing process of the IDS. It includes additional 14 types of new attacks not presented in 10%KDD and in Whole KDD. Thanks to them it is possible to check if the IDS is able to detect a new attack not presented in the training phase.

In the second phase of investigation we added to the KDD 99 data set the simulation results of 14 new attacks (generated with usage of *Metasploit Framework*) and a new normal traffic (instant messaging, VoIP, audio streaming, network games). These types of the network traffic were not presented before in the KDD 99 data set and could show how our proposal works for the real data.

## 4. Proposal of the IDS architecture

The main goal of the IDS architecture proposed in this work is to allow adapting the IDS system to correct classification of a new network traffic related to both: the normal behavior of the end user and to the attacks. The block diagram of the architecture is presented in Fig. 2.

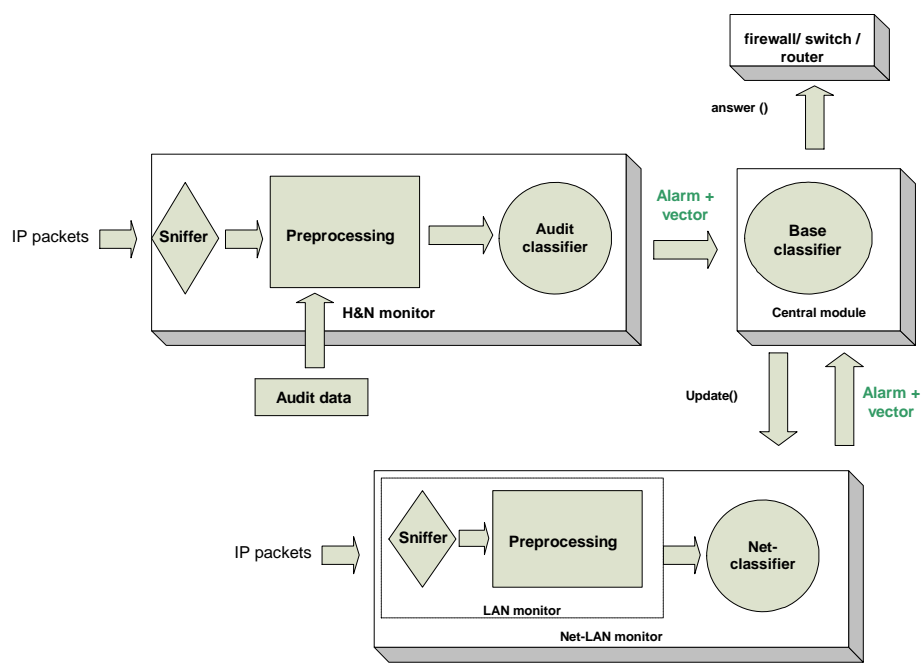

**Fig.2 Block diagram architecture of the IDS proposal**

The IDS system proposed in Fig. 2 is composed of three main modules: *H&N monitor, Net-LAN monitor* and *Central module*. The role of each module is the following:

*H&N monitor*
Main task of this module is analyzing logs and host's audit data in order to find an anomaly event that can be an aggressor's activity. When an attack is detected by *Audit detector*, the network packet associated with it should be identified. Based on that network data, the KDD99 vector of the attack detected is created and is delivered with the alarm flag to *Central module* where it could be used to retrain *Base classifier*.

*Net-LAN monitor*
This module analyses network data provided by a sniffer and transforms it to the KDD99 vector form. The neural network is used for the classification process. It takes a decision whether a current vector is related to an attack or to a normal traffic. In a case of the attack detection, the alarm flag accompanied by the KDD99 vector is sent to *Central module*. Basing on this KDD99 vector it is possible to perform additional analysis and, finally, an eventual decision about classification of this alarm to the attack group or to the false alarm group is taken. For the additional analysis purpose *Central module* can use various methods (in the simplest way, by analysis of a security officer). In a case of the false alarm, the provided KDD99 vector can be used to retrain *Base classifier* and to update all classifiers in all *Net-LAN monitor* modules.

*Central module*

This module obtains all alarms from distributed *Net-LAN* and *H&N monitor* modules and presents them to the end user of the IDS system. The second task of this module is retraining *Base classifier* with using the new learning data provided by *H&N* and *Net-LAN monitors*. *Base classifier* is the neural network that has the same architecture and weights values as *Net-classifier* in each *Net-Lan monitor* module. After the retraining process information about the updated weighs and the new number of hidden neurons is sent to all *Net-LAN monitor* modules.

. An example of the network architecture that can use our solution is shown in Fig. 3.

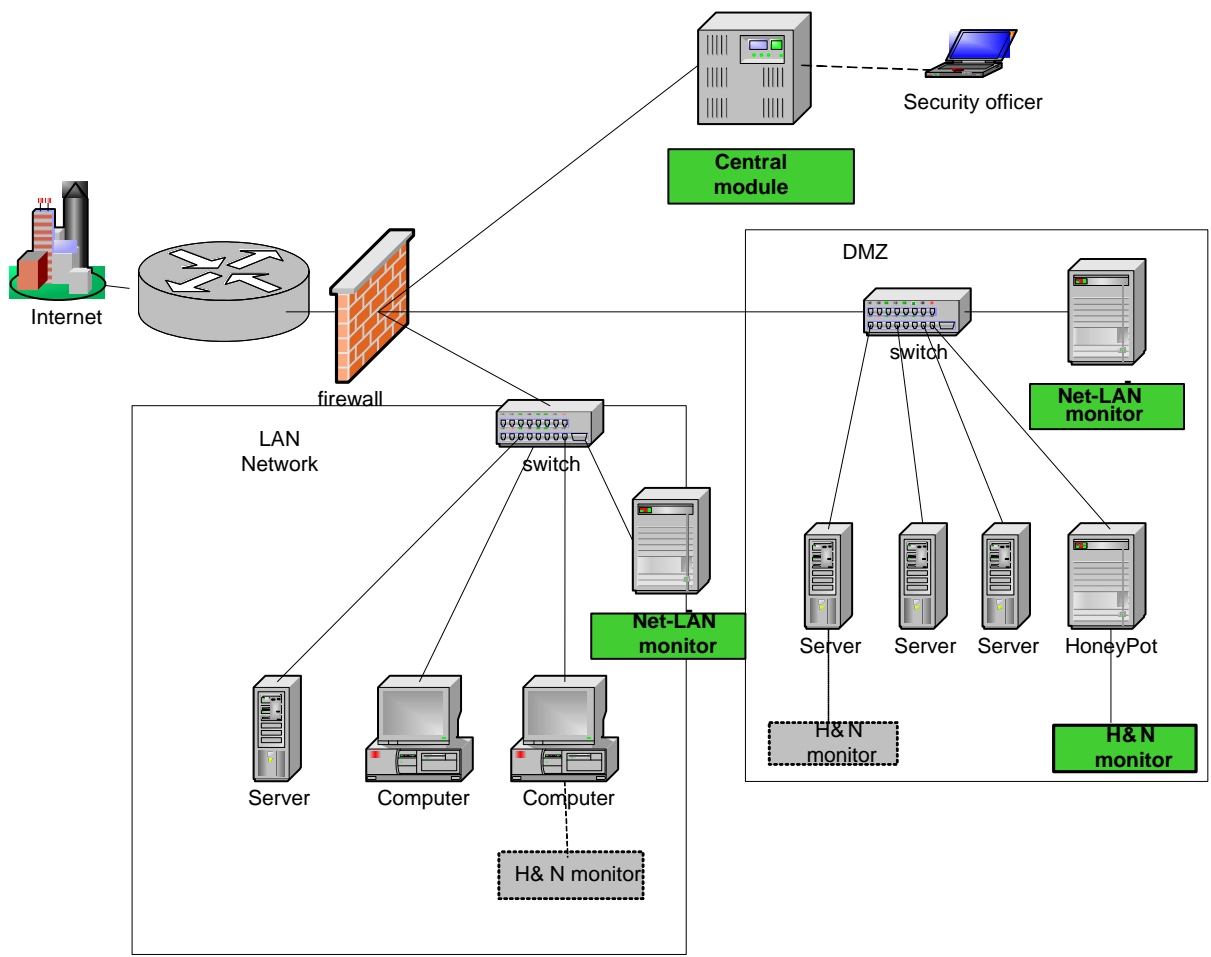

**Fig.3 The network architecture with the IDS proposal**

In our work we decided that *H&N monitor* module is located on a honeypot system. This localization has the following advantages:

- For the honeypot system can be created specific security rules that are less restrictive than the rules for other hosts in the protected network. In that case the host with the installed honeypot system can be visible for an aggressor as an easier target to attack than other hosts. Thanks to this feature it is more probable than for a real production server or for an end user host that the aggressor performs a new attack against it. As a result we can collect the data that represent this new attack. Moreover, focusing an aggressor's attention on the honeypot could distract him from production servers and could increase other hosts safety.
- The honeypot system only simulates some network services and the size of the normal traffic destined for it is very limited. That's why it is easier to identify the data related to the new attacks.

The honeypot system in our proposal is located in the DMZ (*Demilitarized zone*). Thanks to this localization in a case of taking over the honeypot control by an aggressor, the risk of a successful attack against other host located in this internal network is lower because they are protected by specific rules of a firewall. Moreover, the DMZ is less secure so it could attract aggressors and make it easier to collect the data related to the new attacks.

In order to analyze all network traffic data, *Net-LAN monitor* should be located in each real or virtual subnetwork.

## 5. Results of tests

Our investigation was divided in two phases. In the fist phase the accuracy of classification of a new pattern by the neural network was analyzed. The goal of the second phase was to build the prototype of the IDS in the proposed architecture and to check its effectiveness concerning detection of the new attacks and the normal traffic classification. The data sets used for each phase of investigation have been described in Chapter 3.

**Phase 1:**

For the simulation we used two architectures of the neural network: MLP and SVM. For the training and validation the KDD 99 data sets were used. More information about creation and training the neural network can be found in [7].

During analysis of the tests results we noticed that both neural networks architectures have a problem with classification of a new attack not presented in the learning phase.

- For the MLP network the detection rate of a new attack was only 4.26% while for the other attacks presented during the learning phase it was 98 %. The false alarm rate was 2.5%.
- For the SVM network the detection rate of a new attack was only 18.7% while for the other attacks presented during the learning phase it was 97 %. The false alarm rate was 2%.

In Table 2 the accuracy of detection for each type of the new attacks from the test data set was presented. It could be noticed that for 17 new attacks only 2 were classified with the detection rate that equals 100%.

*Table 2. The accuracy of new attacks detection from Corrected KDD testing data set*

| Name of attack | Number of vectors in test data set | Attack detection rate [%] (SVM) | False alarm rate [%] (MLP) |
|---|---|---|---|
| *Snmpgetattack* | 7741 | 0.25 | 0.02 |
| *Named* | 17 | 29.4 | 52.94 |
| *Xlock* | 9 | 0 | 55.55 |
| *Xsnoop* | 4 | 50 | 50 |
| *Sendmail* | 17 | 47.06 | 52.95 |
| *Saint* | 736 | 96.06 | 82.2 |
| *Xterm* | 13 | 84.61 | 61.54 |
| *Mscan* | 1053 | 94.59 | 7.50 |
| *Processtable* | 759 | 77.08 | 1.45 |
| *Ps* | 16 | 37.5 | 81.25 |
| *Apache2* | 794 | 99.75 | 4.16 |
| *Udpstorm* | 2 | 50 | 100 |
| *Httptunnel* | 158 | 16.45 | 0.63 |
| *Worm* | 2 | 0 | 0 |
| *Mailbomb* | 5000 | 6.78 | 0.32 |
| *Sqlattack* | 2 | 100 | 100 |
| *Snmpguess* | 2406 | 0.17 | 0.04 |
| Sum of new attacks | 18729 | 18.7 | 4.26 |

**Phase 2:**

As we noticed in the first phase, the SVM network better classified a new pattern. That's why for the second phase of the investigation we decided to use only this neural network's architecture. The results of the tests are presented in Table 3 (new normal traffic) and Table 4 (new attacks). The first row of each table corresponds to the situation when the new normal traffic and the new attacks were not presented in the learning phase. In that situation for Table 3, false alarms were observed for 4 from 10 new normal vectors analyzed by *Net-LAN monitor* module. For Table 4 it could be noticed that 4 new attacks were not detected. The new attacks not detected by *Net-LAN monitor* were later detected by *H&N monitor* located on the honeypot hosts. Thanks to it we obtain the vectors for retraining the neural networks in *Central module* and for updating *Net-LAN monitor* classifiers. The second row of each table represents situation after this update. In both tables we observe that all the new vectors were properly classified.

*Table 3. The results of Net-LAN monitor classification of the new normal traffic*

| Learning data set (network architecture) | Test data set | Number of false alarms | Number of not detected attacks | Remarks |
|---|---|---|---|---|
| „NaukaIbiza2009sm (SVM)" | „Testowe_gadu3D" | 1347 | 21380 | Falsive alarm concerns: „sip-audio",„rtp-audio", „gadu-k", „radio" |
| „NaukaDay55" | „Testowe_gadu3D" | 1328 | 21432 | All the new normal vectors were properly classified |

*Table 4. The results of detection of the new attacks by Net-LAN monitor*

| Learning data set (network architecture) | Test data set | Number of false alarms | Number of not detected attacks | Remarks |
|---|---|---|---|---|
| „NaukaIbiza2009sm" (SVM) | „Testowe+new31" | 1343 | 21384 | Not detected: *cesarftp_mkd, slimp_ftp_list., mailcarrier_smtp, goodtech_telnet* |
| *net-klasyfikator* after update (SVM) | „Testowe+new31" | 1376 | 21292 | all new attacks were properly detected |

## 6. Conclusions

From our investigation we noticed that the neural networks properly classified the network traffic similar to the one presented during the learning phase. That's why they could be a good solution for detection of the attacks that were modified by an aggressor in order to cheat intrusion detection systems. Unfortunately, the new attacks and the new normal traffic that is significantly different from the one presented in the training phase cannot be classified with sufficiently good accuracy. The IDS architecture that we proposed could improve classification of the new network patterns. Our solution has the following advantages:

- It fixes problems with obtaining training data representing a new attack or a new normal traffic.
- It adapts to new attacks detection.
- Thanks to using *Net- LAN classifier* it is possible to react on an attack in real time and block it before it reaches a host under protection.
- It adapts the system to correct classification of the normal network traffic related to a new service not presented before in the training data (still it should be worked out a method of taking automatically final decision about classification of alarms in *Central module*).

.

It is important to check if adding a new vector influences negatively on the classification accuracy. For example in a situation when we add a new normal traffic to the learning data set and the number of not detected attacks increased significantly, the reason may be that the new vectors can be too much similar to an attack representation. In that case a new feature should be added to KDD99 vectors in order to classify reliably both traffic types.